\documentclass[prl,twocolumn,superscriptaddress,amsmath,amssymb,showpacs]{revtex4}
\usepackage{graphicx}
\usepackage{dcolumn}
\usepackage{bm}
\usepackage{psfig}
\begin{document}
\psfigurepath{./figures}
\title{Magnetic transition and magnetic structure of Sr$_4$Ru$_3$O$_{10}$}

\author{Wei Bao}
\affiliation{Los Alamos National Laboratory, Los Alamos, New Mexico 87545}
\author{Z.Q. Mao}
\author{M. Zhou}
\author{J. Hooper}
\affiliation{Department of Physics, Tulane University, New Orleans, Louisiana 70118}
\author{J.W. Lynn}
\affiliation{NIST Center for Neutron Research, National Institute of Standards
and Technology, Gaithersburg, Maryland 20899}
\author{R.S. Freitas}
\author{P. Schiffer}
\author{Y. Liu}
\affiliation{Department of Physics, Pennsylvania State University, University Park, Pennsylvania 16802}
\author{H.Q. Yuan}
\author{M. Salamon}
\affiliation{Department of Physics, University of Illinois at Urbana-Champaign, Urbana, Illinois 61801}

\date{\today}

\begin{abstract}
We have investigated the magnetic transition and magnetic structure
of triple-layered ruthenate Sr$_4$Ru$_3$O$_{10}$ directly using
neutron scattering techniques. Only one ferromagnetic phase is
observed, and previously proposed antiferromagnetic phase
transitions are ruled out. The complex anisotropic magnetotransport,
magnetization and in-plane metamagnetic behaviors of this quasi
two-dimensional (2D) material are most likely due to magnetic domain
processes with strong magnetocrystalline anisotropy and a strongly
anisotropic demagnetization factor.
\end{abstract}


\maketitle

Strontium ruthenates of the Ruddlesden-Popper (RP) series,
Sr$_{n+1}$Ru$_n$O$_{3n+1}$, exhibit a rich variety of fascinating
physical properties, such as unusual spin-triplet superconductivity
in Sr$_2$RuO$_4$ ($n$=1)
\cite{Ishida_214_Nature,Hayden_214_PRL,Mackenzie_214_RMP,Nelson_214_Science},
metamagnetic quantum criticality in Sr$_3$Ru$_2$O$_7$ ($n$=2)
\cite{Perry_327_PRL,Grigera_327_Science,Grigera_327PhaseForm_Science},
and itinerant ferromagnetism in SrRuO$_3$ ($n$=$\infty$)
\cite{jpsj_113a,Allen_113}. They have played a pivotal role in the
past decade in the study of novel physics of strongly correlated
electron systems. Sr$_4$Ru$_3$O$_{10}$ is the $n$=3 member of the RP
series with triple layers of corner shared RuO$_6$ octahedra
separated by SrO rock-salt double layers. While Sr$_2$RuO$_4$ and
Sr$_3$Ru$_2$O$_7$ are close to magnetic instabilities
\cite{Sidis_incom,Singh_327,Capogna_327}, bulk magnetization studies
of Sr$_4$Ru$_3$O$_{10}$ provide evidence for true long-range
ferromagnetism with $T_C \approx 105$ K
\cite{stru4310,Cao_4310_PRB}, lower than that seen in SrRuO$_3$
($T_C \approx 160$ K) \cite{jpsj_113a,Allen_113}. The ferromagnetism
in this material exhibits very unusual anisotropic properties under
applied magnetic fields. Magnetization curves are distinctly
different for magnetic fields applied along the $c$ axis and in the
basal plane, and an additional transition at $T^*\approx 50$ K has
been indicated \cite{stru4310,Cao_4310_PRB}. Below $T^*$, a
metamagnetic transition is induced by a magnetic field applied in
the plane. Very unusual transport properties, including ultrasharp
magnetoresistivity steps and a non-metallic temperature dependence
of resistivity, have been observed near this transition
\cite{Mao_PhaseSep4310_PRL}. In addition, strong magnetoelastic
coupling has been observed in this
material.\cite{Gupta_Raman4310_PRL}

It has been proposed that in Sr$_4$Ru$_3$O$_{10}$ the Ru magnetic
moments are canted; the $c$-axis components order in a ferromagnetic
transition at $T_C\approx 105$~K, while the in-plane components
order in a separate antiferromagnetic transition at $T^*\approx 50$
K \cite{Cao_4310_PRB,Gupta_Raman4310_PRL}. Within this picture, the
metamagnetic transition is thought to be a transition from an
antiferromagnetic to a ferromagnetic arrangement for the in-plane
moment components. We have performed direct measurements of the
magnetic structure in Sr$_4$Ru$_3$O$_{10}$ by neutron scattering,
which yields a result that is inconsistent with this proposed
magnetic picture. We find only a ferromagnetic transition at $T_C =
100$ K with a normal order parameter and with an easy axis lying in
the $ab$ plane. This observation, together with magnetotransport and
magnetization measurements, strongly suggests that the metamagnetic
behavior for $H$//$ab$ results from strong magnetocrystalline
anisotropy and ferromagnetic domain formation. Our result clarifies
the metamagnetic behavior in the ferromagnetic state of
Sr$_4$Ru$_3$O$_{10}$, which has been a longstanding puzzle. This
finding is of fundamental importance in understanding other novel
properties in this class of materials
\cite{Gupta_Raman4310_PRL,Mao_PhaseSep4310_PRL}.

Single crystals of Sr$_4$Ru$_3$O$_{10}$ used for this study were
grown using a floating zone technique. The high quality of the
crystals is indicated by the small residual resistivity in the 1.5-6.0
$\mu\Omega\cdot$cm range
\cite{Mao_PhaseSep4310_PRL}.
The tiny orthorhombic distortion of the
$Pbam$ crystal structure \cite{stru4310} is negligible for this
work, and we use the $I4/mmm$ tetragonal unit cell to label the
reciprocal space with dimensions of $a=b=3.895$ and $c=28.46$ \AA \
at 70~K. Samples used in this neutron scattering work at NIST are
thin plates with the shortest dimension along the $c$-axis. A 0.1 g
sample was oriented to investigate the ($hhl$) zone in reciprocal
space and a 0.4 g sample for the ($h0l$) zone. A small amount of
SrRuO$_3$ intergrowth was observed in our experiments,
which can be easily separated by its very
different lattice parameter $c=7.86 \AA$.
Both triple-axis and
two-axis modes of the thermal triple-axis spectrometer BT9 were used
with neutrons of energy $E=14.7$~meV. The collimations were 40-48-44
in the two-axis mode, and 40-48-44-80 in the triple-axis mode.
Pyrolytic graphite filters of 10 cm total thickness were used to
reduce higher order neutrons. Sample temperature was regulated by a
closed cycle refrigerator in zero field experiments, and by a 7 T
superconducting cryomagnet in magnetic field experiments.

\begin{figure}[t]
\vskip -3ex
\centerline{\psfig{file=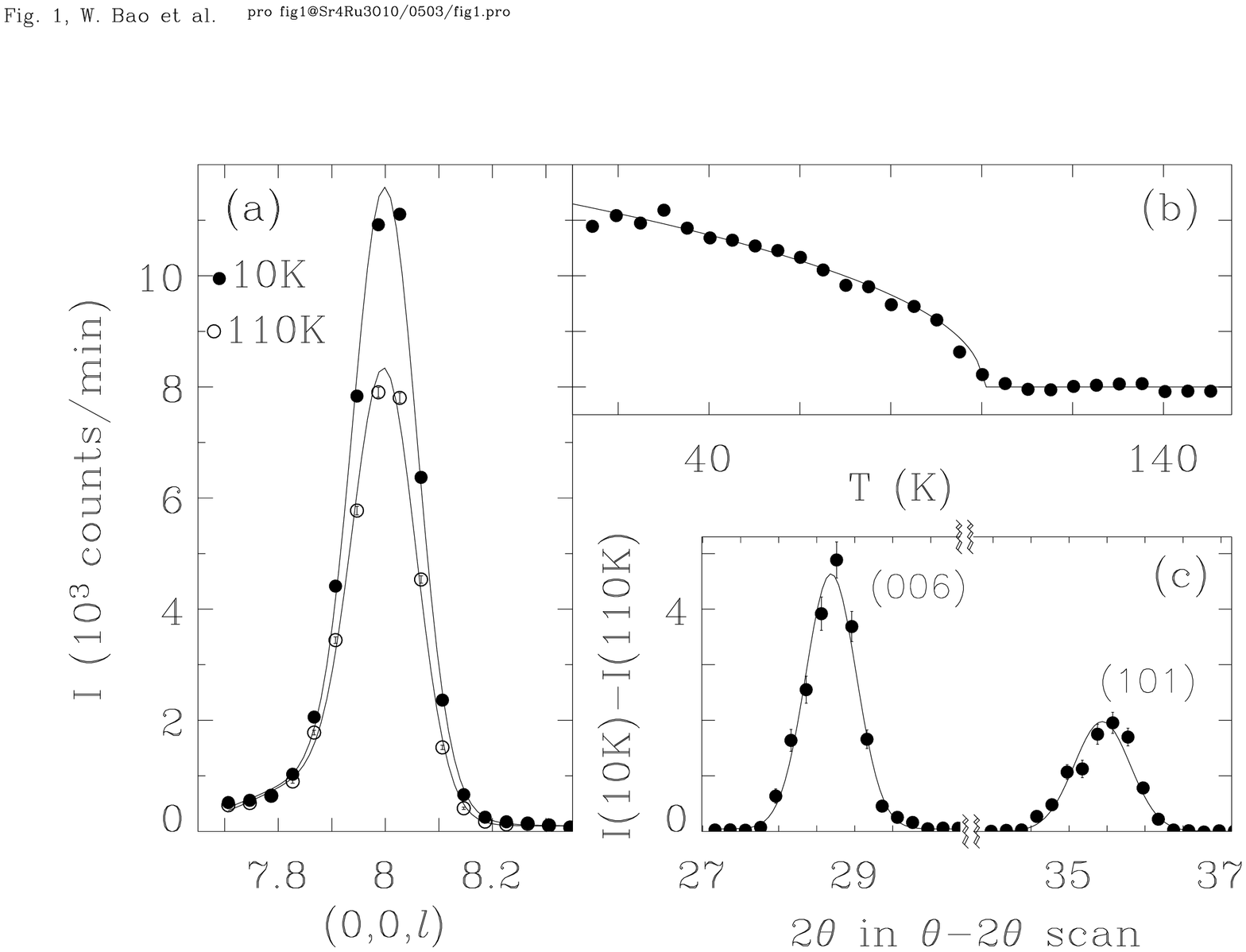,width=.9\columnwidth,angle=90,clip=}}
\vskip -3ex
\caption{Magnetic Bragg peaks coincide with structural Bragg peaks
in Sr$_4$Ru$_3$O$_{10}$. (a) Scans along the interlayer direction
through the (008) Bragg peak at 10 and 110 K. (b) Temperature
dependence of the (008) Bragg peak, showing the magnetic transition at
$T_C=100$ K. (c) Difference scans in the same unit as (a),
representing {\em magnetic} Bragg intensities at (006) and (101).
}\label{fig1}
\end{figure}

Fig.~\ref{fig1}(a) shows the (008) Bragg peak of
Sr$_4$Ru$_3$O$_{10}$ measured at 10 and 110~K. The enhanced
intensity at low temperature is due to the magnetic transition at
$T_C=100$~K. The resolution-limited peak width remains the same at
low temperature and above $T_C$, indicating stacking of the RuO$_6$
and SrO layers according to the crystal structure to form a
macroscopic crystal, and a magnetic transition in the whole crystal.
Scans, covering an area larger than a Brillouin zone, in both the
($hhl$) and ($h0l$) reciprocal zones, detected no additional
temperature-dependent magnetic Bragg peaks, either commensurate or
incommensurate. Hence, the magnetic structure in
Sr$_4$Ru$_3$O$_{10}$ follows the same symmetry of the crystal
structure. Antiferromagnetic structures, such as layers of Ru
moments stacking alternately along the c axis
\cite{Gupta_Raman4310_PRL}, are forbidden.

Fig.~\ref{fig1}(b) shows the temperature dependence of the (008)
Bragg peak, which serves as the squared magnetic order parameter of
the phase transition. The order parameter behaves normally below
$T_C=100$ K. In particular, it does not exhibit either extra enhancement at
$T^*\approx 50$~K as in the $c$-axis magnetization measurement, or a
sharp drop below $T^*$ as in the in-plane magnetization
\cite{stru4310,Cao_4310_PRB}. Thus, the anomalies around 50~K in
bulk magnetization measurements are not likely due to another
magnetic transition in Sr$_4$Ru$_3$O$_{10}$, which is consistent
with the fact that the specific heat of this material shows no
anomaly at 50~K \cite{Brill_specificH,Jo_Specific heat_cond}.

\begin{figure}[t]
\vskip -3ex
\centerline{\psfig{file=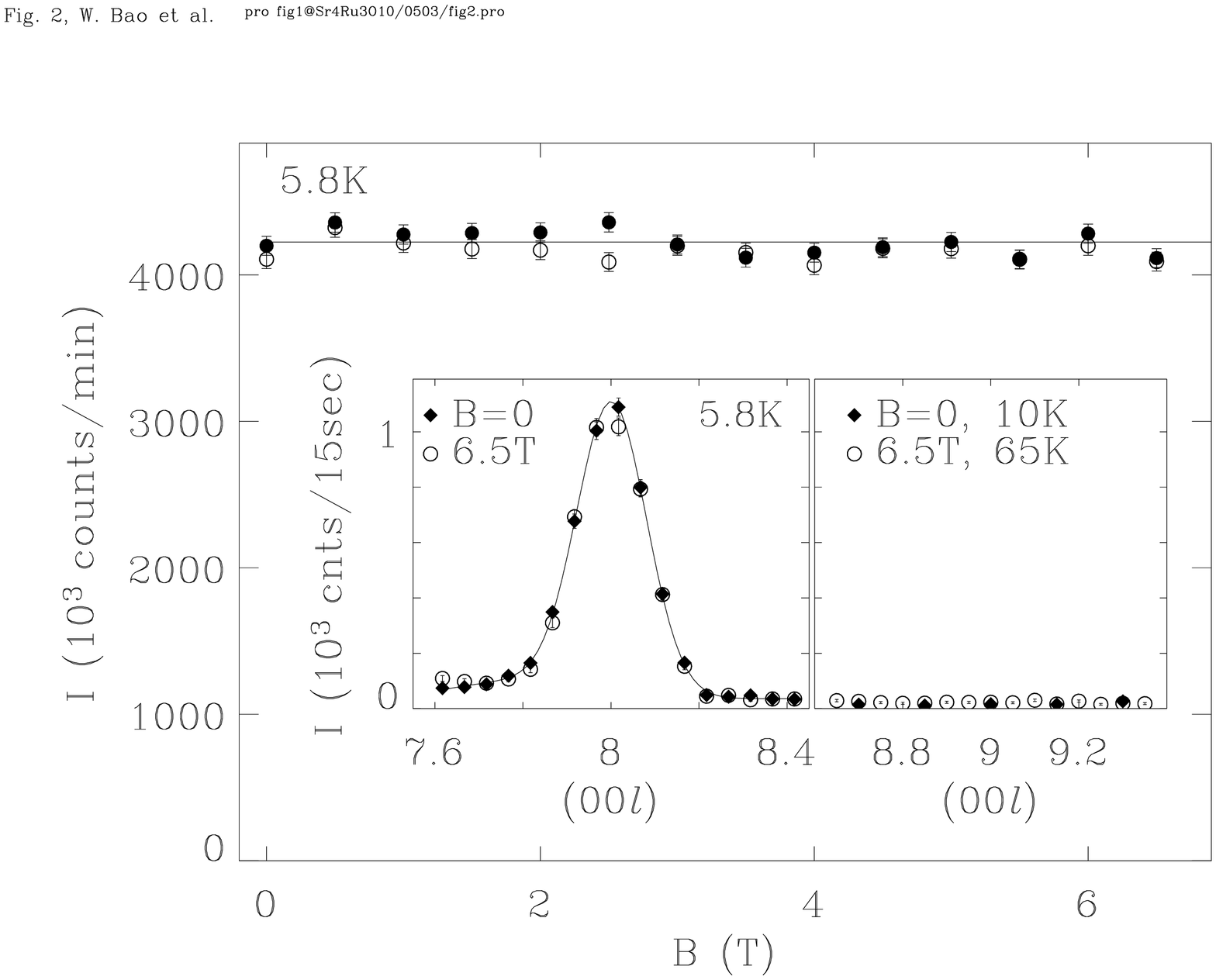,width=.9\columnwidth,angle=90,clip=}}
\vskip -3ex
\caption{Dependence of the (008) Bragg peak at 5.8 K on magnetic
field applied along the in-plane [1\={1}0] direction. The solid
circles are for field ramping up, and open circles for ramping down.
Inset: Scans through the allowed (008) or the forbidden (009) peak
at B=0 (solid diamonds) and 6.5 T (open circles). }\label{fig2}
\end{figure}

Although the proposed antiferromagnetic arrangement is excluded by
the selection rule for magnetic Bragg peaks, it is still interesting
to investigate the magnetic order parameter under magnetic field.
The 0.1 g sample was mounted in a cryomagnet with magnetic field
along the [1\={1}0] direction. Fig.~\ref{fig2} shows the (008) Bragg
intensity at 5.8 K with magnetic field ramping up (solid circles)
and down (open circles). No first-order metamagnetic transition like
that in the magnetization \cite{Cao_4310_PRB} was observed. The Bragg
peak profile below and above $B_c$ also remains the same, as shown for
scans at 5.8 K for B=0 (solid symbols) and 6.5 T (open symbols) in
the inset to Fig.~\ref{fig2}. Similarly, no field effect can be
detected in either the profile or intensity of the (008) Bragg peak
at 18, 40 and 65 K. The magnetic field also does not induce extra
intensity at forbidden peaks, such as (009), (see inset to
Fig.~\ref{fig2}).

The neutron scattering cross section from a magnetic structure
is given by \cite{neut_squire}
\begin{equation}
\sigma({\bf q})=\left(\frac{\gamma r_0}{2}\right)^2
    \left|f(q)\right|^2
    \sum_{\mu,\nu}(\delta_{\mu\nu}
    -\widehat{\rm q}_{\mu}\widehat{\rm q}_{\nu})
    {\cal F}^*_{\mu}({\bf q}){\cal F}_{\nu}({\bf q}),
\label{eq_cs}
\end{equation}
where $(\gamma r_0/2)^2=0.07265$~barns/$\mu_B^2$, $f(q)$ is the
atomic form factor of the Ru ion \cite{Hayden_214_PRL,Sidis_incom},
$\widehat{\bf q}$ the unit vector of ${\bf q}$, and ${\cal
F}_{\mu}({\bf q})$ the $\mu$th Cartesian component of the magnetic
structure factor per Sr$_4$Ru$_3$O$_{10}$. The factor
$(\delta_{\mu\nu}-\widehat{\rm q}_{\mu}\widehat{\rm q}_{\nu})$ in
Eq.~(\ref{eq_cs}) dictates that only magnetic moment components
which are perpendicular to the neutron wave-vector transfer, {\bf
q}, contribute to magnetic neutron scattering.
In particular, if the Ru magnetic moments actually pointed along the c-axis then the magnetic intensities would vanish at Bragg points (00$l$) where $l$ is an even integer;  thus magnetic intensity at peaks such as (008)
in Fig.~\ref{fig1} and \ref{fig2} rules out this proposed ferromagnetic structure \cite{stru4310,Cao_4310_PRB}. Furthermore, the
field independent (008) Bragg intensity as shown in Fig.~\ref{fig2}
rules out a canted magnetic structure with a $c$-axis component
which rotates to the basal plane for $B>B_c$\cite{Gupta_Raman4310_PRL}.
Instead, magnetic moments are aligned in
the basal plane, and thus the rotation of the moments to the
(1\={1}0) direction when the field $B>B_c$ would not change the neutron
scattering cross section. This conclusion is consistent with the easy axis of
the related compound SrRuO$_3$, which lies in the basal plane
\cite{jpsj_113a,jpsj_113b}.  In addition, the demagnetization factor
of the plate-like Sr$_4$Ru$_3$O$_{10}$ samples also favors the
in-plane moment orientation\cite{old_mag}.

It is well known for ferromagnetic materials that the bulk magnetization
does not directly reflect the magnetic order parameter due to domain formation
when the magnetic field is below a certain strength \cite{old_mag}.
Complex anisotropic behaviors in magnetic field  can be induced
by domain processes due to strong magnetocrystalline anisotropy,
demagnetization factor, magnetostatic energy, magnetostriction and mutual interaction between the magnetization
and the external field\cite{old_mag,jpsj_113c}.
For example, SrRuO$_3$ has a very strong magnetocrystalline anisotropy.
The {\em internal} crystalline anisotropy field is about 2T for the
pseudo-cubic SrRuO$_3$, with easy axis along the [110] directions
\cite{jpsj_113a}. In the pseudo-tetragonal SrRuO$_3$, the easy axis
lies along [100], while the $c$-axis is a direction of medium magnetization with
the spin-flop field at 1.5 T \cite{jpsj_113b}. Substantial magnetic field
is required to line up magnetic moments of domains.
Since the local environment for Ru in
Sr$_4$Ru$_3$O$_{10}$ is similar to that in SrRuO$_3$, it is not
unrealistic to expect a similarly strong magnetocrystalline
anisotropy in Sr$_4$Ru$_3$O$_{10}$, and complex magnetic behaviors
can be understood within such a domain picture.
\begin{figure}[b]
\includegraphics[width=0.3\textwidth,]{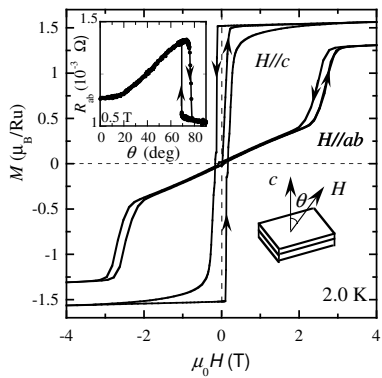}
\caption{Magnetization loop for Sr$_{4}$Ru$_{3}$O$_{10}$ for
in-plane and out-of-plane field orientations. Inset at lower right:
schematic of field rotation with respect to the c-axis for the
electrical transport measurement. Inset at upper left: the  polar
angle dependence of the in-plane resistance. All samples used in this work
are from the same batch. }\label{fig3}
\end{figure}

In Fig.~\ref{fig3}, we present the magnetization loop for
Sr$_4$Ru$_3$O$_{10}$ for in-plane and out-of-plane field
orientations. Consistent with the previously reported results
\cite{stru4310,Cao_4310_PRB}, the magnetization curve is extremely
anisotropic. For $H//c$, the remnant magnetization
is almost equal to the saturated magnetization and the
magnetization curve shows discontinuous jumps in the second and
fourth quadrants, in sharp contrast with zero remnant magnetization
for $H//ab$. The out-of-plane field leads to behavior
consistent with typical ferromagnets, and the larger magnetization
is tempting for assigning the $c$-axis as the easy axis
of magnetization. However, our neutron experiment demonstrates
that the easy axis of Sr$_4$Ru$_3$O$_{10}$ lies in the $ab$-plane.
In addition, the in-plane field behavior is atypical for a ferromagnet
and resembles a metamagnetic transition at $\sim$2.5 T.
But our neutron scattering experiment (see Fig.~\ref{fig2}) does not
detect any increase in magnetic order parameter near 2.5 T.
Thus, the metamagnetic behavior for $H//ab$ cannot be interpreted
as the Stoner transition as in Sr$_{3}$Ru$_{2}$O$_{7}$
\cite{Perry_327_PRL,Grigera_327_Science,Grigera_327PhaseForm_Science}.

Magnetic domains form in Sr$_4$Ru$_3$O$_{10}$, like in any sizable ferromagnet. Cancellation of magnetization of domains with different
in-plane easy axes is responsible for the low magnetization at low
in-plane field, and a large field of $~2.5$ T is required to
induce spin-flip and spin-flop in this material of strong magnetocrystalline anisotropy, as in SrRuO$_{3}$ \cite{jpsj_113c}.
The energy barrier between the $c$-axis and an easy axis is likely much lower than
that between two easy axes in the basal plane, which explains
the easier alignment of moments along the c-axis.
This domain picture is also
consistent with the angular dependence of in-plane
magnetoresistance of Sr$_{4}$Ru$_{3}$O$_{10}$, see insets in Fig.~\ref{fig3}.
When the 0.5 T applied field rotates from the $ab$-plane ($\theta=\pi/2$)
toward the $c$ axis ($\theta$=0), the magnetoresistance exhibits a
sharp jump at $\sim$20 degree away from the $ab$-plane, indicating
a first-order process in which the in-plane magnetization vectors
of magnetic domains flop to the $c$-axis.

Now let us turn to the determination of {\em microscopic} magnetic structure
of Sr$_4$Ru$_3$O$_{10}$.
The same symmetry for the magnetic structure as the crystalline one requires
ferromagnetic alignment of magnetic moments within a Ru layer. Thus
the exchange interactions between moments within the plane should be
ferromagnetic, instead of antiferromagnetic as suggested in the
Raman work \cite{Gupta_Raman4310_PRL}. The same symmetry requirement
also dictates identical magnetic arrangements in every
RuO$_6$ triple-layer building block. With the moment oriented in the
plane, what remains to be determined for the magnetic structure,
then, is the relative magnitude and angle of magnetic moments in the
three neighboring layers of a building block. The relation between
magnetic moments in the two outer layers are further constrained by
the symmetry. Therefore, the magnetic moments at the $a$ site of the
$I4/mmm$ or $Pbam$ space-group in the center plane and at the $e$ sites of the two
outer planes can be generally written as
\begin{subequations} \begin{gather}
{\bf M}_{e1}= \alpha M [ \widehat{\bf x}\cos(\beta) + \widehat{\bf y}\sin(\beta) ] \\
{\bf M}_{a}= M \widehat{\bf x} \\
{\bf M}_{e2}= \alpha M [\widehat{\bf x}\cos(\beta) - \widehat{\bf y}\sin(\beta) ],
\end{gather} \label{eq_M}
\end{subequations}
where $M$ is the moment size at the $a$ site, $\alpha$ a number, $\beta$ the angle
between moments in the nearest neighbor planes, and
$\widehat{\bf x}$ and $\widehat{\bf y}$ are mutually perpendicular
unit vectors in the basal plane. The average moment, as measured by a
magnetometer, is $M_{ave}=M [1+2\alpha \cos(\beta)]/3$, provided that the applied
magnetic field does not alter $\beta$.

To determine the free parameters in Eq.~(\ref{eq_M}), a set of
magnetic Bragg intensities was measured. A couple of examples are
shown in Fig.~\ref{fig1}. The
magnetic cross sections of Sr$_4$Ru$_3$O$_{10}$ can be conveniently determined
through the relation,
\begin{equation}
\sigma_{obs}({\bf q})=\frac{I(10K)-I(110K)}{I(110K)} \, |F_S({\bf q})|^{2} ,
\label{eq_ms}
\end{equation}
where $F_S({\bf q})$ is the known structure factor per Sr$_4$Ru$_3$O$_{10}$
of the structural Bragg peak
at {\bf q} \cite{stru4310}. These are listed in Table~I.
\begin{table}
\caption{Magnetic Bragg cross-section, $\sigma_{obs}$,
observed at
10~K in barns per Sr$_4$Ru$_3$O$_{10}$.
The theoretical cross-section, $\sigma_{I}$,
in the same units, is calculated for the collinear
magnetic structure using $\alpha=0.44$, $\beta=0$ and $M=1.85(2) \mu_B$/Ru  in
Eq.~(\ref{eq_M}). $\sigma_{II}$ is calculated for the non-collinear
magnetic structure using $\alpha=0.60$, $\beta=29^{\circ}$ and $M=1.59(1) \mu_B$/Ru  in
Eq.~(\ref{eq_M}). $|F_S|^2$ is the squared structure factor
for the crystal structure in barns per Sr$_4$Ru$_3$O$_{10}$.
}
\label{mlist}
\begin{ruledtabular}
\begin{tabular}{clccc}
${\bf q}$ & $\sigma_{obs}$ & $\sigma_{I}$ &  $\sigma_{II}$ & $|F_S|^2$ \\
\hline
     ( 0      0      2   )&   0.64(1) & 0.66 & 0.65 & 1.564 \\
     ( 0      0      4   )&   0.11(4) & 0.03 &0.036 & 2.850\\
     ( 0      0      6   )&   1.11(5) & 1.33 &1.36  & 10.85\\
     ( 0      0      8   )&   1.69(8) & 1.12 &1.15 & 3.945\\
     ( 0      0      10  )&   0.023(2) & 0.022 &0.022 & 0.207\\
     ( 1      0      1   )&   1.06(6) & 0.59 & 0.60 & 8.771\\
     ( 1      0      3   )&   0.02(1) & 0.014 & 0.015 & 1.294\\
     ( 1      0      5   )&   3.2(4) & 0.13 &0.12 & 2.342\\
     ( 1      0      7   )&   5(2) & 0.66 & 0.68 & 195.1\\
     ( 1      0      9   )&   0.06(2) & 0.12 &0.12 & 1.359\\
    (1 1 2) & 0.00(4) & 0.08 & 0.08 & 1.927\\
    (1 1 4) & 0.00(5) & 0.004 & 0.005 & 6.678
\end{tabular}
\end{ruledtabular}
\end{table}
For (103), (105) and (107), the scattered neutron beam was nearly parallel
to the sample plate, and the measured $I(110K)$ cannot be properly accounted for
by the squared structure factor $|F_S|^2$ due to extinction/absorption.
Therefore, $\sigma_{obs}$ for them in Table I
are not reliable. Among the remaining Bragg peaks, (002), (006), (008), (0,0,10)
and (101) gain more than 10\% intensity at low temperature due to the ferromagnetic transition.

The simplest magnetic structure consistent with Eq.~(\ref{eq_M})
is the one with identical magnetic moments for the Ru ions, namely,
$(\alpha, \beta)=(1,0)$ in Eq.~(\ref{eq_M}), or
${\bf M}_{e1}={\bf M}_{a}={\bf M}_{e2}$.
For this magnetic structure, Eq.~(\ref{eq_cs}) is reduced to
\begin{equation}
\sigma({\bf q})=\left(\frac{\gamma r_0 f(q)}{2}\right)^2
    \frac{1+\cos^2(\beta)}{2} \left|{\cal F}_{x}({\bf q})\right|^2,
\label{eq_cs1}
\end{equation}
where $\beta$ is the angle between {\bf q} and the $c$ axis, and
equal populations of magnetic domains are assumed for the zero field
experiments. However, Eq.~(\ref{eq_cs1}) does not provide a good
description of the measured $\sigma_{obs}$. The $\chi^2$ can be
reduced by a order of magnitude in a least-squares fit of
$\sigma_{obs}$ to Eq.~(\ref{eq_cs}) with unrestricted  $\alpha$ and
$\beta$ for the magnetic model in Eq.~(\ref{eq_M}). Both $(\alpha,
\beta)_I=(0.44, 0)$ and (0.6, 29$^{\circ})_{II}$ with
$\chi^2\approx 21$ best fit the data, and an arc connecting them in
the $\alpha$-$\beta$ plane offer a slightly inferior fit. Hence,
there is no unique result. In Table I we present theoretical cross
sections $\sigma_I$ for the collinear magnetic structure $(\alpha,
\beta)=(0.44, 0)$ with $M=1.85(2)$ $\mu_B$/Ru, and $\sigma_{II}$
for the non-collinear structure (0.6, 29$^{\circ}$) with $M=1.59(1)$
$\mu_B$/Ru. Using a subset of $\sigma_{obs}$ in Table I with
$I(10K)/I(110K) >1.1$, or including data of larger uncertainty in
the least-square fit does not affect the results. The average
moments of the two magnetic models are $M_{ave}=1.2$ and 1.1
$\mu_B$/Ru respectively, both comparable to the experimental value
shown in Fig.~\ref{fig3}.

In summary, we found only one magnetic transition at 100 K with easy
axis in the basal plane for Sr$_4$Ru$_3$O$_{10}$.
No additional antiferromagnetic transitions are observed.
Our experiments strongly suggest that the complex magnetotransport,
magnetization and metamagnetic behaviors results from magnetic domain
processes in this quasi 2D material of strong magnetocrystalline anisotropy.

Work at LANL were supported by US DOE, at Tulane by the Louisiana
Board of Regents support fund LEQSF(2003-06)-RD-A-26 and pilot fund
NSF/LEQSF(2005)-pfund-23 and a grant from Research Corporation.

\bibliographystyle{unsrt}

\end{document}